\documentclass[3p]{elsarticle}
\usepackage{amssymb}

\usepackage{graphics}
\usepackage{subfigure}
\usepackage{epstopdf}
\usepackage{amsmath}

\journal{Astroparticle Physics}

\begin{document}

\begin{frontmatter}

\title{Prospects for the detection of GRBs with HAWC}

\author[label1]{I. Taboada}
\cortext[corresponding]{Corresponding author: itaboada@gatech.edu}
\address[label1]{School of Physics and Center for Relativistic Astrophysics, Georgia Institute of Technology, Atlanta, GA USA }
\author[label2]{R.C. Gilmore}
\address[label2]{Santa Cruz Institute for Particle Physics. University of California. Santa Cruz, CA USA.}

\begin{abstract}
The observation of Gamma Ray Bursts (GRBs) with very-high-energy (VHE)
gamma rays can provide understanding of the particle acceleration
mechanisms in GRBs, and can also be used to probe the extra-galactic
background light and place constraints on Lorentz invariance
violation. We present prospects for GRB detection by the ground-based
HAWC (High Altitude Water Cherenkov) gamma-ray observatory. We model the VHE spectrum of
GRBs by extrapolating observations by Fermi LAT and other
observatories to higher energies. Under the assumption that only
e-pair production associated with extra-galactic background light is
responsible for high-energy cutoffs in the spectrum, we find that HAWC
will have a detection rate as high as 1.65 GRBs/year. Most of the
sensitivity of HAWC to GRBs is derived from short-hard GRBs during the
prompt phase. We explore the possibility of universal high-energy
cutoffs in GRB spectra and find that the GRB detection rate by HAWC
should be at least half of this figure as long as the typical
intrinsic cutoff is above 200-300 GeV in the rest frame. 
\end{abstract}


\end{frontmatter}


\section{Introduction}
\label{sec:intro}

Gamma Ray Bursts (GRBs) are among the most powerful events in the
Universe, and have been the subject of observational studies from
radio to multi-GeV energies. Fermi LAT has shown that GRBs are able to
produce photons up to observed energies of 94 GeV ($\approx$126~GeV
after redshift correction) for GRB 130427A \cite{grb130427a}. It is unknown up to what
energy the spectrum extends, as present-day observations are limited
by effective area, in the case of space-based instruments, and by
slewing constraints and energy threshold for ground-based Imaging Air
Cherenkov Telescopes. Studying the spectrum beyond 10 GeV is of interest in
understanding GRB mechanisms themselves, and also allows us to probe
cosmological phenomena such as the extra-galactic background light
(EBL) and it may be used to constrain Lorentz invariance violation. 
The GRB proper, or prompt emission, is detected in
the keV-MeV band by space-borne instruments such as Fermi-GBM
\cite{FermiGBM} and Swift-BAT \cite{SwiftBAT}. There are at least two known classes of GRBs, short-hard
GRBs and soft-long GRBs. The short-hard variety (sGRBs) have durations
of less than 2~s and a peak of $\nu F_\nu$ at Earth is at 
$\sim$1~MeV. SGRBs are believed to be the result of the merger of two
compact objects \cite{janka,rosswog}, such as two neutron stars or a neutron star and a
solar mass black hole. Long-soft GRBs (LGRBs) last more than 2~s and
the peak of $\nu F_\nu$ is at $\sim$50~keV. Core collapse supernova
type Ic are believed to the progenitors of LGRBs and coincidences of
the two have been observed \cite{grb030329}. For both types of GRBs,
emission is believed to happen in highly relativistic narrow jets that
point in the direction of Earth.

In this paper we will show that the GRB detection rate by the extended
air shower array detector HAWC may be as high as 1.65 GRBs per year,
assuming that the spectrum is only cutoff by EBL attenuation, and that
short GRB detection over the lifetime of the instrument is likely as
long as intrinsic spectral cutoffs are not typical below $\approx$150 GeV in
the rest frame. The most important energy range for detection by HAWC
is in 50 GeV - 500 GeV range.


Satellites with instruments sensitive to hard
gamma-rays, such as CGRO and Fermi, have extended the
observations from 30~MeV to tens of GeV. GRB~130427A \cite{grb130427a} that was observed up to 94 GeV, or 126~GeV once corrected for
redshift, shows that GRBs are capable of producing very-high-energy (VHE)
photons. As of May 2013, seven GRBs have been observed above 10 GeV
\cite{grb080916c,grb090510,grb090902b,grb090926a,FermiLATgrbCatalog,grb130427a}. 


The keV-MeV burst spectrum is usually well described by the Band
function \cite{band}. At higher energies, Fermi LAT has shown that additional 
non-Band components are needed for every single high signal to noise
ratio GRB \cite{FermiLATgrbCatalog}. This
additional component can be a hard power law with no obvious high
energy cutoff, such as GRB~090510 \cite{grb090510}. It can also be a cutoff
in the high energy extension of the Band function such as
GRB~100724B \cite{FermiLATgrbCatalog}. Or it can be a hard power law with a high energy
cutoff, such as GRB~090926 \cite{FermiLATgrbCatalog}. For LGRBs, all cases of additional
components have been observed in high signal/noise GRBs, but for sGRBs, only an additional power
law with no cutoff has been observed. The number of bursts with high signal to noise
ratio is low \cite{FermiLATgrbCatalog}, thus, the relative frequency of each
type of additional component is fairly uncertain. 
%

Satellites are ideal platforms for GRB observations above 100~MeV 
and below a few tens of GeV. However, because of their limited size,
and because spectra fall off as a power law of energy, they are not appropriate to extend
the observations to even higher energies. The $>$30~GeV study of the non-Band components is of interest in
several respects. A cutoff in the spectrum may be indicative of 
e-pair production at the source and thus a measurement of the Lorentz boost factor
of the jet.
A cutoff is also possible due to EBL attenuation. For redshifts lower
than $z\approx$1, the EBL cutoff energy is $\gtrsim$100~GeV and
probably would not be observable
by Fermi-LAT or other space borne instruments. Observation of the $>$30~GeV spectrum may also be used
to constrain Lorentz invariance violation and to model particle
acceleration in GRBs. Finally spectral and temporal information of
$>$30~GeV emission may help constrain GRB models.


Ground based detectors are ideal
for $>$30~GeV studies of GRBs. Two techniques are available: Air Cherenkov Telescopes
(IACTs) and Extended Air Shower arrays (EAS). IACTs have better
sensitivity both due to larger effective area and better background
rejection. However IACTs have limited field of view, which requires
them to slew fast after an alert to observe GRBs. The best slewing time for
VERITAS is $\approx$100~s \cite{VeritasGRB}, which results in observations that are in the
early afterglow phase. The other disadvantage is that IACTs operate
only in good weather, at night, with no or limited moon light. The
duty cycle of IACTs is 10-15\% and bright GRBs, that may result in
$>$30~GeV observations are rare events. Finally, IACTs need GRBs to have a small localization error,
so as guarantee that the search area fits into the field of view. EAS
arrays have lower sensitivity, but the have the advantage of very high
duty cycle (over 95\%) and very large instantaneous field of 
view (nominally 2~sr) so they are able to catch rare bright
GRBs. Furthermore, the wide field of view allows EAS arrays to study
GRBs with a localization error of several degrees, as is often the
case for GRBs reported by Fermi GBM. 

A previous study \cite{CTAgrbRate} of the prospects for the detection of
GRBs by CTA, a proposed 3$^{\mathrm{rd}}$ generation IACT, found that
CTA will be mostly sensitive to LGRBs in the afterglow phase, with a
detection rate of 0.5-2 GRBs/yr (depending on detector configuration,
GRB satellite alert rate, and high energy spectral model applied)
assuming that the only cutoff in the high-energy spectrum is due to
EBL \cite{CTAgrbRate}. In the current publication we will follow a
similar method as used for CTA, but applied to HAWC.

\section{HAWC}
\label{sec:hawc}

The HAWC (High Altitude Water Cherenkov) observatory is an EAS array
in construction in Central Mexico at 4100~m 
asl. Once completed it will consist of 300 water tanks of 7.3~m in
diameter and 4.5~m of depth.
Each tank will contain three 8" PMTs and
one high quantum efficiency 10" PMT. HAWC is sensitive to GRBs with
two modes of operation, the \textit{main DAQ} and the \textit{scaler
  DAQ}. The main DAQ operates by measuring the arrival time of the
particle shower front product of the interaction of a high energy gamma
ray in the upper atmosphere. Angular resolution for energies relevant to GRBs is
$\approx 1^\circ$ and energy resolution is very poor, as the detector
operates near the threshold \cite{HAWCgrb}. The scaler DAQ operates
in the single particle technique \cite{Vernetto}, in which a large
transient fluence of gamma rays is
found as a statistical excess in the count rate of all PMTs in the
detector combined. The scaler DAQ does not provide localization and
has not energy resolution. The energy response of both the main DAQ and scaler
DAQ are different, thus the spectrum of a jointly observed GRB may be
constrained \cite{HAWCgrb}. Once operational HAWC will be the EAS
array with the best sensitivity to GRBs and would be able to detect
historical GRBs, such as 090902B and 090510 if they fall in its field
of view \cite{HAWCgrb}. 

The publication by HAWC that describes
the sensitivity to GRBs \cite{HAWCgrb}, excludes the central 10''  PMT.
The results
presented here use the effective area presented in the HAWC publication. 
We interpolate both in zenith and energy to obtain the effective
area for any zenith in the range 0 to $40^\circ$ and any energy
from 12 GeV to 1 TeV for the main DAQ. For the scaler DAQ, the
zenith range is the same but energy is interpolated in
the range 1~GeV to 1~TeV, reflecting better sensitivity to softer
spectra provided by the scaler system. We use the background
parameterization of both the main DAQ and the scaler DAQ presented in
the HAWC publication \cite{HAWCgrb}. For the main DAQ we use a
trigger threshold of 30 PMTs, since the HAWC collaboration has
indicated it believes it can operate at a very low threshold
\cite{HAWCgrb}. 

As suggested in the HAWC publication we use a search circle of
$1.1^\circ$ for the main DAQ. As we indicated before, HAWC and other
EAS arrays can study GRBs with poor localization. For GRBs reported by Fermi GBM alone,
the localization uncertainty is $\approx10^\circ$. HAWC can tile the
the sky with $\approx 80$ search regions, corresponding to applying a search trials factor of $\approx 
80$. In our simulation of sGRBs we find that the background in 2
seconds for the main DAQ ranges from 2 to 16 events with an average of
7.5 events. The background rate in our calculation is exclusively a
function of zenith angle. The number of counts needed for a
5$\sigma$ excess over a expected 7.5 background events is 26 counts -
or 18.5 signal events (the p-value for 26 counts is $2.9 \times
10^{-8}$, but 25 counts is below the 5$\sigma$ threshold). With a
trials factor of 80, the number of counts needed would be 28, or 
20.5 signal counts. Tiling the sky results in a loss of 
sensitivity of $\approx$10\% with respect to knowing the GRB location
with high precision. This simplistic calculation shows that we can ignore the
trial factors for tiling the sky in searching for GRBs with poor
localization. It should be further noted that bright Fermi GBM bursts
have typically better localization than dim ones and as the
simulations that we present here show, GRBs that can be detected by
HAWC are expected to be bright as observed by Fermi GBM.

The sensitivity of HAWC to GRBs is in part a function of the time
window in which VHE emission is searched. The shorter the
burst the better sensitivity as the background is reduced. In our
calculations here we simplistically assume that the search will be
performed in a window of width $T_{90}$ as measured by Fermi GBM for
LGRBs and in a fixed window of 2~s for sGRBs.

\section{Modeling of VHE emission by GRBs}
\label{sec:model}

We have built a model for $>$1~GeV spectrum for GRBs and simulated a
large number of pseudo-GRBs. We assume that HAWC observations are triggered by Fermi
GBM. While there are other satellites
that contribute to a significant number of GRB alerts each year, notably Swift
BAT, we find that bright GRBs are the most likely to result in
detections by HAWC. Thus it is more important for HAWC to follow a
very wide field of view space borne detector such as Fermi GBM,
instead of a more sensitive detector such are Swift BAT.
The study of CTA GRB detection rate found that most detections would
be triggered by Swift because, as opposed to Fermi GBM, it provides
localizations of an arcminute or better. 

We model the $>1$~GeV emission assuming an additional power law
component that extends up to 1~TeV (at Earth) and is cutoff
exclusively by EBL. The simulation samples from fits to the
distribution of redshift for both LGRBs and sGRBs. 
The spectrum above 1~GeV is assumed to be a power law. We normalize
the power law given a relationship found between the fluences measured
by Fermi GBM and Fermi LAT. 
Correlations among parameters that describe GRBs are not simulated.


The simulated spectrum is folded with HAWC's interpolated effective
area. The sensitivity for each GRB is found taking into account the
parameterized background. To account for
other potential sources of cutoff other than EBL we complement the
calculation described above with a similar one that adds a universal
Heaviside cutoff in the GRB self frame ranging from 100 GeV to 500
GeV. 
 

\subsection{$>1$~GeV spectrum}

Observations by Fermi LAT and Fermi GBM show a correlation between the low
energy fluence, reported in the 10 keV - 10 MeV band (henceforth the
GBM band), and the high energy fluence, reported in the 100 MeV - 10 GeV 
band (henceforth the LAT band). There is significant
scatter around this correlation, and in the case of sGRBs, only 2
GRBs, 081024B and 090510, provide a hint of the correlation
\cite{FermiLATgrbCatalog}. The duration ($T_{90}$) as measured in the
two energy bands is different, and the LAT observation is usually delayed
with respect to the GBM observation. 
However the correlation of fluences remains true even if the LAT
observations are restricted to GBM's $T_{90}$. Indeed, approximately half of the fluence in the LAT band
is produced in coincidence with GBM's $T_{90}$ \cite{FermiLATgrbCatalog}. 
For LGRBs, the LAT fluence in GBM's $T_{90}$ is approximately 10\% the GBM fluence and for sGRBs this ratio
is over 100\% \cite{FermiLATgrbCatalog}. 
A critical ingredient in our model is the selection of 100\% (10\%) fluence ratio for sGRBS (LGRBs).

\begin{figure}[ht]
  \centering
  \subfigure
  {  
    \includegraphics[width=.45\textwidth]{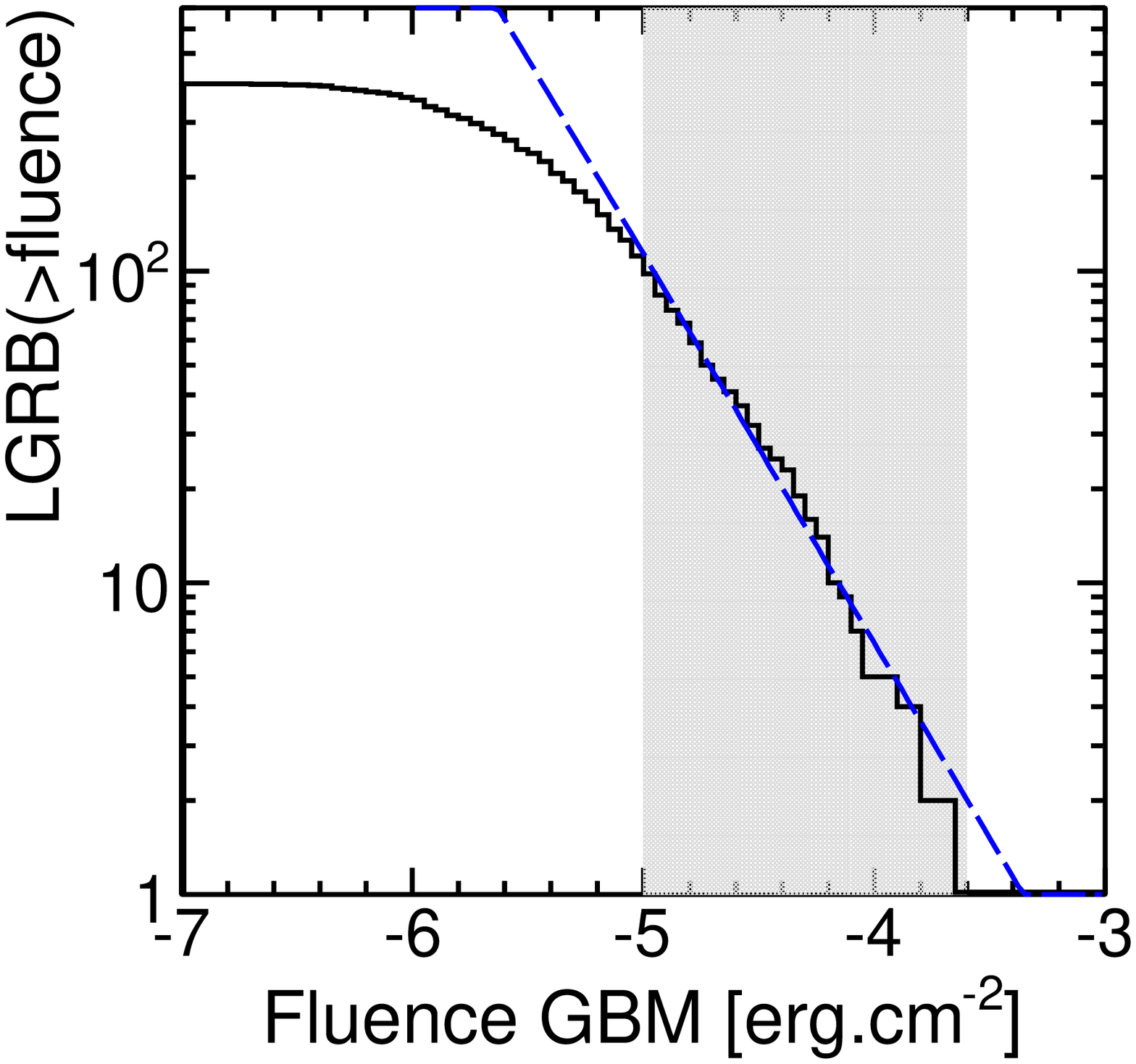}
  }
  \subfigure
  {
    \includegraphics[width=.45\textwidth]{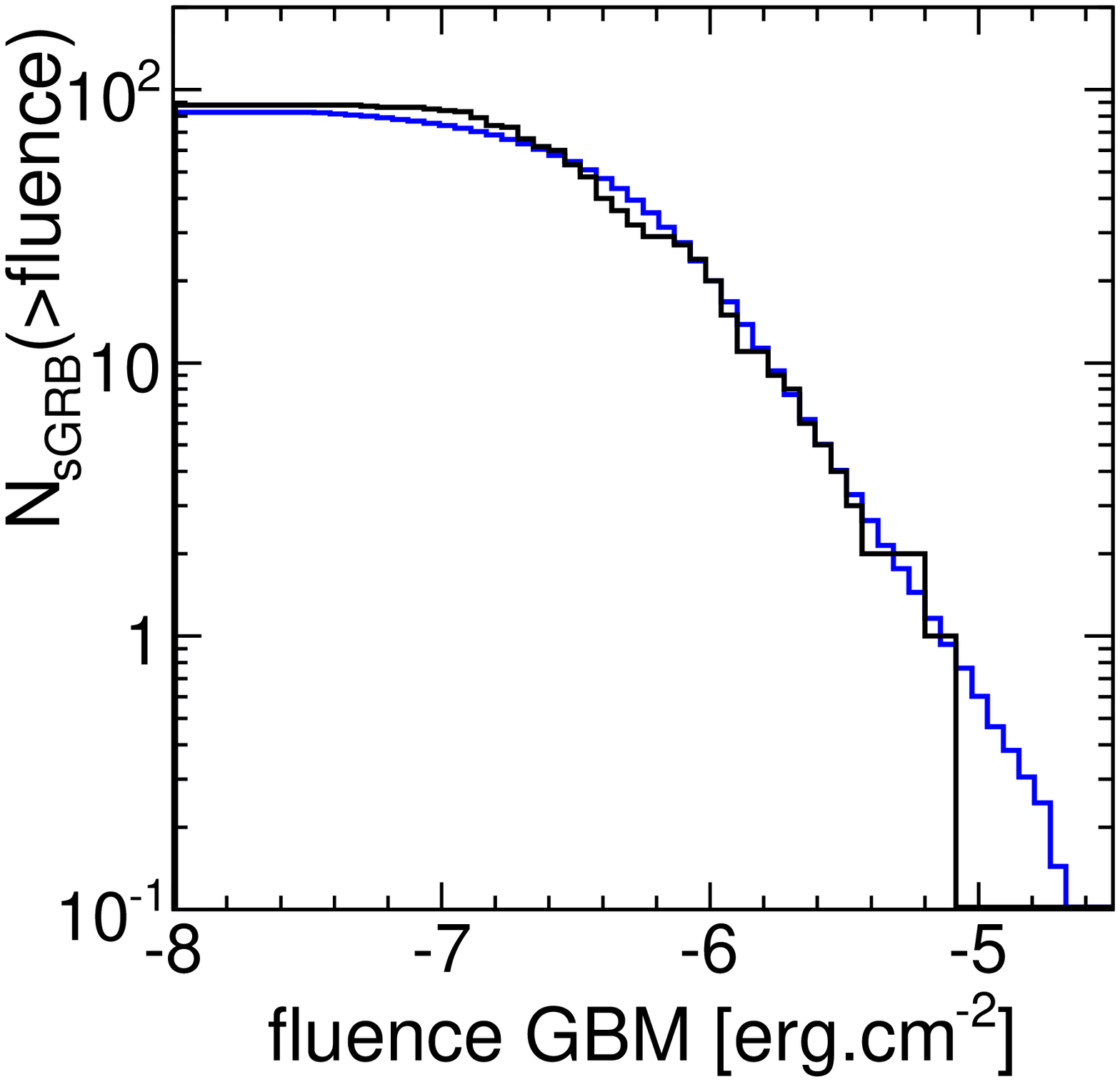}
  }
  \caption{\label{fig:GBMfluence} Cumulative distribution of GRBs
   as a function of fluence measured by Fermi GBM in the 10~keV -
   10~MeV band \cite{FermiGBMcatalog}. The distribution of LGRBs
   (left) has been fitted with a power    law using the shaded
   region. For sGRBS (right) a more elaborate 
   function has been fitted to the data. At high fluence the
   cumulative distribution of sGRBs is also well described by a power
   law.
}
\end{figure}

We have parameterized Fermi GBM's fluence using the data from the 2
year GBM catalog \cite{FermiGBMcatalog}.  Fig. \ref{fig:GBMfluence}
shows that the cumulative distribution of bright GRBs are well
described by a power law.  At low fluences the effect of GBM's trigger
becomes important. In our model, bright GBM GRBs are those that result
in HAWC observations, thus a very detailed description of Fermi GBM's
trigger is not critical. The Fermi GBM trigger would naturally
introduce a correlation between GBM fluence and GBM $T_{90}$. As HAWC
will observe GRBs mostly above the GBM trigger, we can ignore this
correlation. 

For LGRBs we have fitted the distribution of the number of GRBs above a given GBM fluence in
the range $1 \times 10^{-5}$ - $2.5\times 10^{-3}$ erg$\cdot$cm$^{-2}$. We find
that the LGRB distribution is well described by a power law of index
-1.5. This index is consistent with previous observations
including those by Fermi GBM \cite{FermiGBMcatalog}. We normalize the rate of LGRBs in Fermi GBM by observing that
there are 29 LGRBs in the 2 year GBM catalog above fluence of $3
\times 10^{-5}$~erg$\cdot$cm$^{-2}$. Figure \ref{fig:GBMfluence} shows the distribution of
GBM fluences for LGRBs as well as the power law fit. For fluences
below $\approx 1\times10^{-5}$~erg$\cdot$cm$^{-2}$, the power law does not account for
trigger effects of Fermi GBM. A result of the calculations that we
present here is that only 3\% of the GRBs detectable by HAWC would
have a GBM fluence below $1\times10^{-5}$~erg$\cdot$cm$^{-2}$.

We have followed a similar procedure to parameterize the fluence of
sGRBs. For sGRBs that are well above the GBM threshold, we also
find that a power law of index -1.5 also provides a good
description. However we find a larger fraction of sGRBs with fluence near
the GBM threshold can result a detection by HAWC. A simple threshold
on $F_{GBM}/\sqrt{T_{90}}$, where $F_{GBM}$ is the GBM
fluence does not accurately describe the detector threshold. The GBM
trigger probably is better described in terms of the peak flux from a
sGRB, and thus it depends on the details of the lightcurve of the 
sGRBs. To avoid unnecessary complexity we instead parameterize the
cumulative distribution of sGRBs above a
certain GBM fluence including an arbitrary function to represent the
trigger and we ignore the
correlation with GBM's $T_{90}$. This is a valid approach as the
calculations presented here assume that HAWC performs its search in a
fixed 2~s window for sGRBs. Figure \ref{fig:GBMfluence} shows the
cumulative sGRBs above a given GBM fluence along with the cumulative
distribution of simulated pseudo-GRBs that follow the fitted function. To normalize the absolute rate of sGRBs in Fermi
GBM, we note that in the 2 year Fermi GBM catalog \cite{FermiGBMcatalog} there are 20 sGRBs above a fluence of $1
\times 10^{-6}$~erg$\cdot$cm$^{-2}$. 

For both LGRBs and sGRBs we have tested values of the index that
differ from -1.5 to describe the cumulative rate of GRBs as a
function of GBM fluence. Using a value of -1.25 results in a detection
rate that changes by only 3\%.

We assume that every single simulated pseudo-GRB,
of both long-soft and short-hard type, has an additional power law
component. This is already in contradiction with observations by LAT
of high signal to noise GRBs \cite{FermiLATgrbCatalog}. At the very
minimum it is in contradiction for LGRBs. 
Therefore our present calculations for LGRBs are an over
estimate of the rate in HAWC. There are only a handful of high signal
to noise ratio GRBs in LAT data, and these are not clearly identified
as such in the 3 year Fermi LAT
GRB catalog \cite{FermiLATgrbCatalog}. However we estimate that one third to one half of the high signal to noise ratio
LAT LGRBs have an additional power law with no indication of cutoff in
the LAT band. For LGRBs with an additional
power law in LAT, the central value of the spectral index is -2 so we
assume this value in our simulations. 

Only 2 sGRBs have high signal to noise ratio in LAT, GRB~081024B
\cite{grb081024b} and 090510 \cite{FermiLATgrbCatalog}. They both
have a very hard additional component of index -1.6.
%
Evidence for additional power law components have
been seen on other GRBs for which Fermi LAT observations are not
available. The three brightest sGRBs in GBM, GRB~090510, 090227B
and 090228,  all have additional power law components, in the
10~keV-10~MeV range, with very hard indices \cite{guiriec}.
Fermi GBM can't by itself determine if
there is a high energy cutoff for GRB~090227B and 090228. However it is clear that additional
power laws are common in bright sGRBs. In our simulations we will
assume that all sGRBs have an additional power law without cutoff with
spectral index -1.6, consistent with evidence for sGRBs mentioned above.

The spectral index for the additional power law of LGRBs and sGRBs has
the largest effect in our calculation. This is because we are
extrapolating observations by Fermi LAT in the 0.1-10 GeV range up to
hundreds of GeV.

\subsection{Redshift distribution}

\begin{figure}[ht]
  \centering
  \subfigure
  {  
    \includegraphics[width=.45\textwidth]{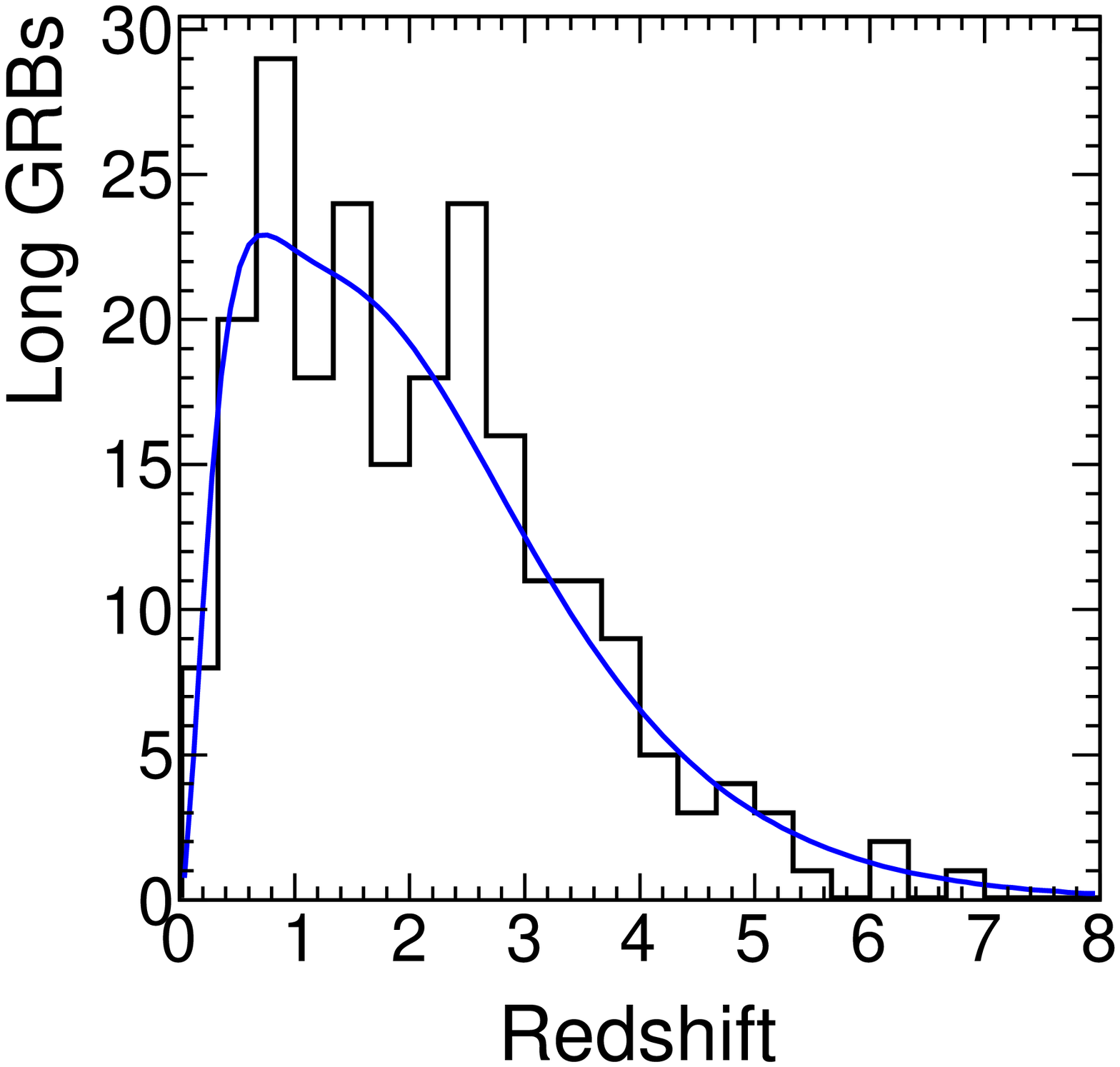}
  }
  \subfigure
  {
    \includegraphics[width=.45\textwidth]{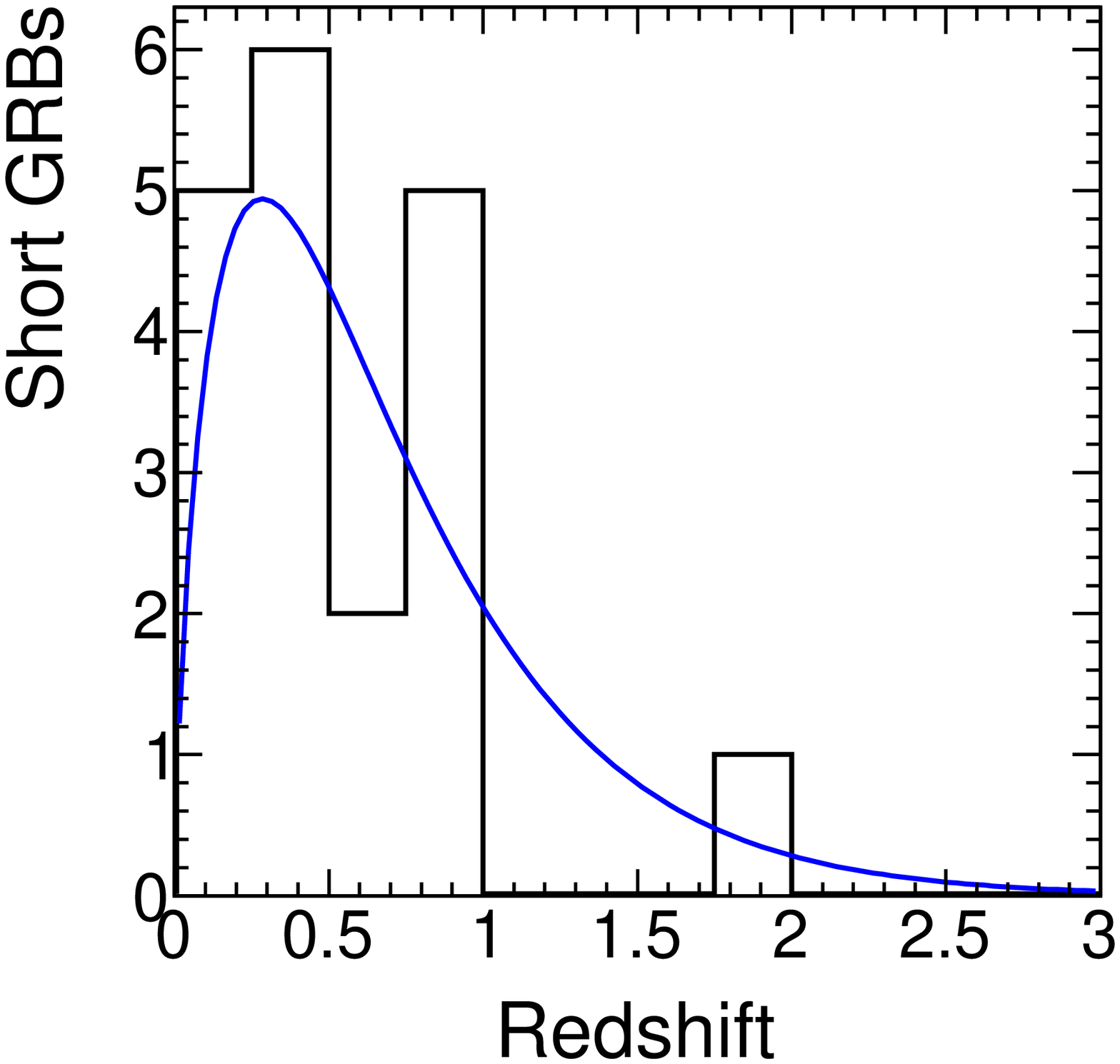}
  }
  \caption{\label{fig:redshift} Redshift distribution for long (left)
    and short (right) GRBs. The curve shown is a function fitted to
  the data. LGRB data was collected from the online Swift catalog
\cite{SwiftData}  See appendix for details on the
  selection of sGRBs. }
\end{figure}

E-pair production of the VHE GRB spectrum with
extra-galactic background light results in a cutoff of the
spectrum. This effect is redshift dependent, with higher redshifts
resulting in a lower cutoff value. In our simulations we have
parameterized the redshift distribution for both LGRBs and sGRBs. 

For LGRBs we have used bursts of $T_{90}>2$~s in the online Swift catalog
\cite{SwiftData}. The list of LGRBs was collected in March 2013. 
We exclude GRBs of $z \le 0.1$ as they may be a
separate population of very low luminosity GRBs that we are not trying
to describe here. We find 223 LGRBs. The redshift distribution is fitted with an ad hoc
function shown in figure \ref{fig:redshift}. This is the very same
function used for the calculation of the GRB rate in CTA \cite{CTAgrbRate}, but fit
parameters are different, as more LGRBs were used in the fit presented here.

Using a combination of the online Swift catalog \cite{SwiftData} and other sources, we find 19 sGRBs with
redshift measurements. An ad hoc function has been used to fit the
distribution of sGRBs and is shown in figure \ref{fig:redshift}. The
list of sGRBs with redshift that we have used may be found in the appendix.

In our model we simulate GRBs of $z\ge 0.1$ and extend the spectrum at
Earth up to a maximum of 1~TeV. These two choices are consistent as
1~TeV is approximately the cutoff expected for $z=0.1$. Because the rate
of GRBs of both types is very low below z=0.1, this choice of minimum
redshift and maximum energy of the spectrum does not have a
significant impact in our calculations. To calculate the effect of EBL
attenuation and cutoff we use the 2009 model by Gilmore et
al. \cite{Gilmore09}, which is based on a semi-analytic treatment of
galaxy formation and evolution and predicts a UV-optical EBL flux
level similar to that seen in recent observationally-motivated models
and galaxy survey data.

\subsection{$T_{90}$ distribution for LGRBs}

\begin{figure}[ht]
  \centering
  \includegraphics[width=.45\textwidth]{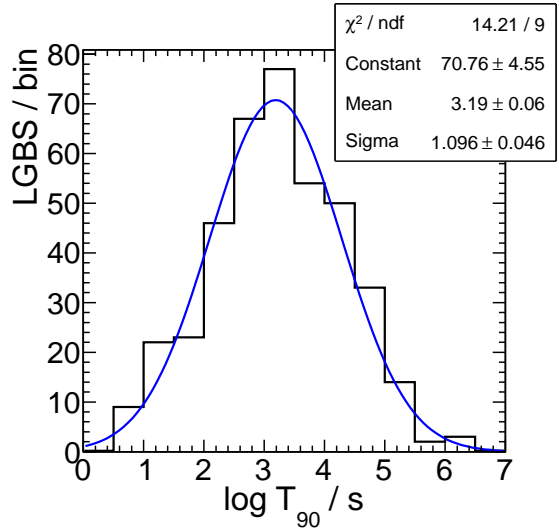}
  \caption{\label{fig:T90} Distribution of $T_{90}$ as measured for
    Fermi GBM in the 2 year catalog \cite{FermiGBMcatalog}. The fit is a log-normal
    distribution. A $\chi^2$ of 14.2 with 9 degrees of freedom has a
    p-value of 0.115.}
\end{figure}

As described in section \textsection~\ref{sec:hawc}, we will assume that HAWC
searches for VHE emission in a window of width equal to
$T_{90}$ as measured by GBM. We have parameterized the LGRB duration as
using Fermi GBM's 2 year catalog \cite{FermiGBMcatalog}. The
distribution of $T_{90}$ is well described by a log-normal function. Figure
\ref{fig:T90} shows the data from the Fermi GBM 2 year catalog and the
fit we have used. 

\section{Detection rate in HAWC}

\begin{figure}[ht]
  \centering
  \subfigure
  {  
    \includegraphics[width=.31\textwidth]{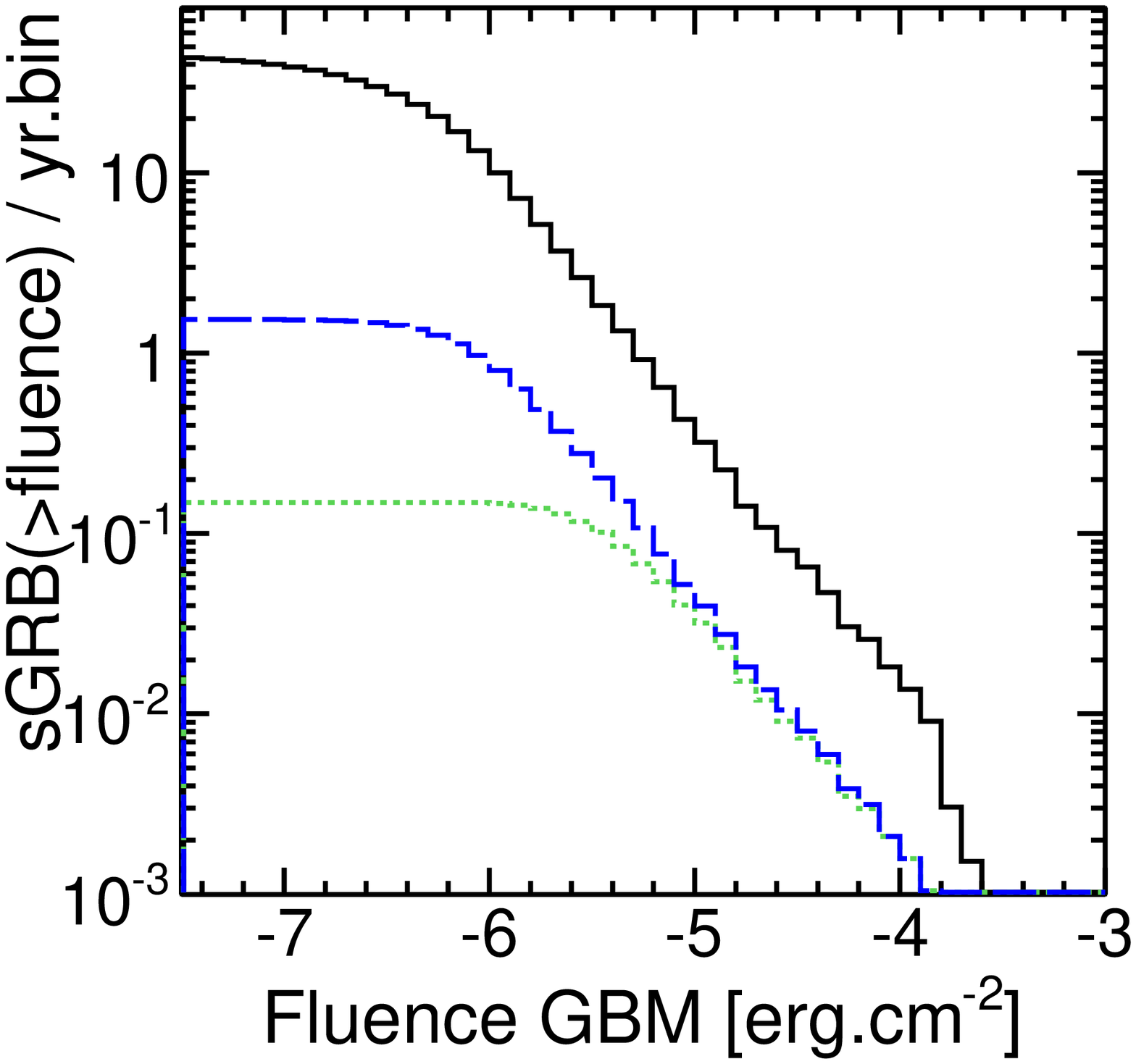}
  }
  \subfigure
  {
    \includegraphics[width=.31\textwidth]{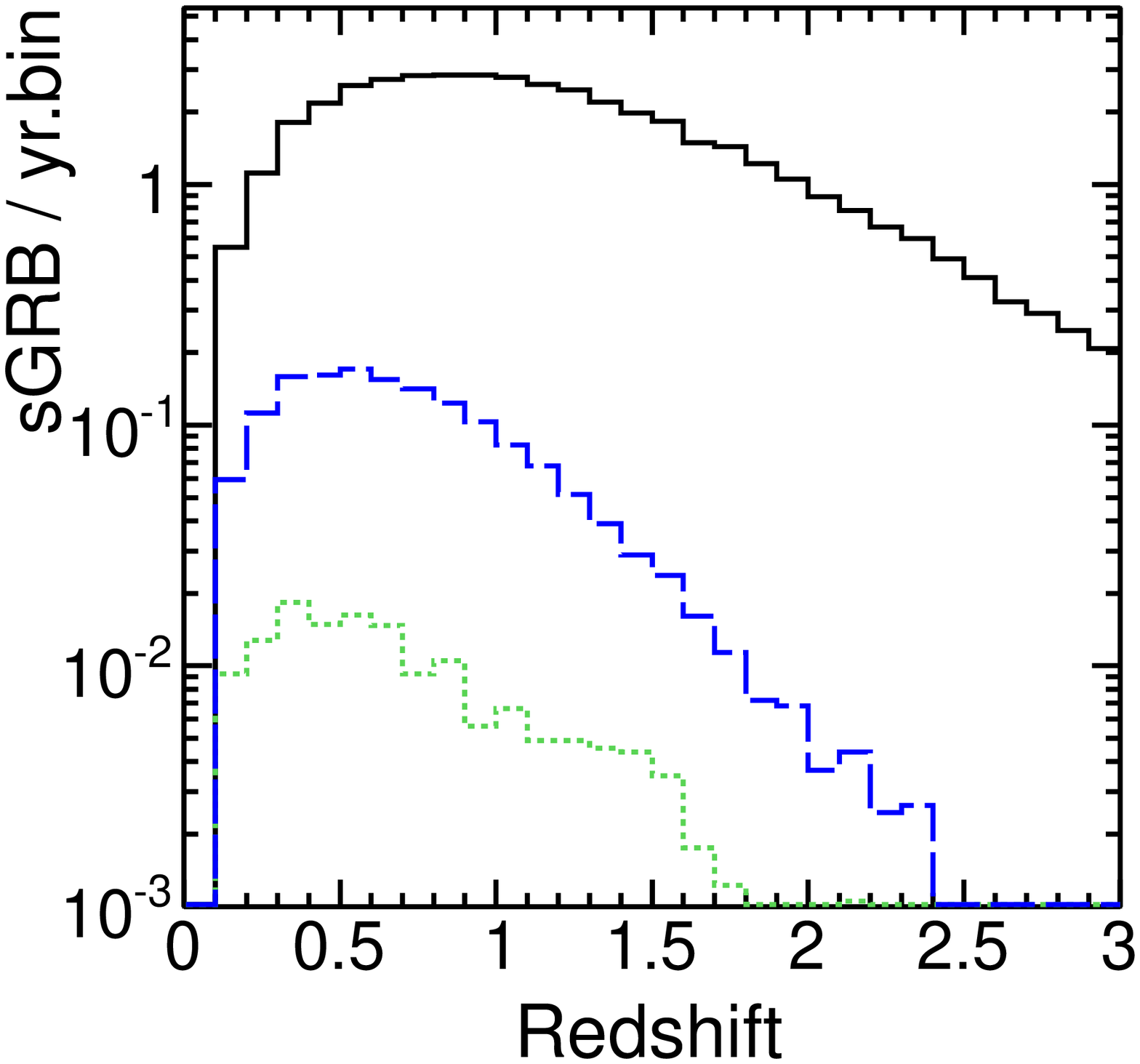}
  }
  \subfigure
  {
    \includegraphics[width=.31\textwidth]{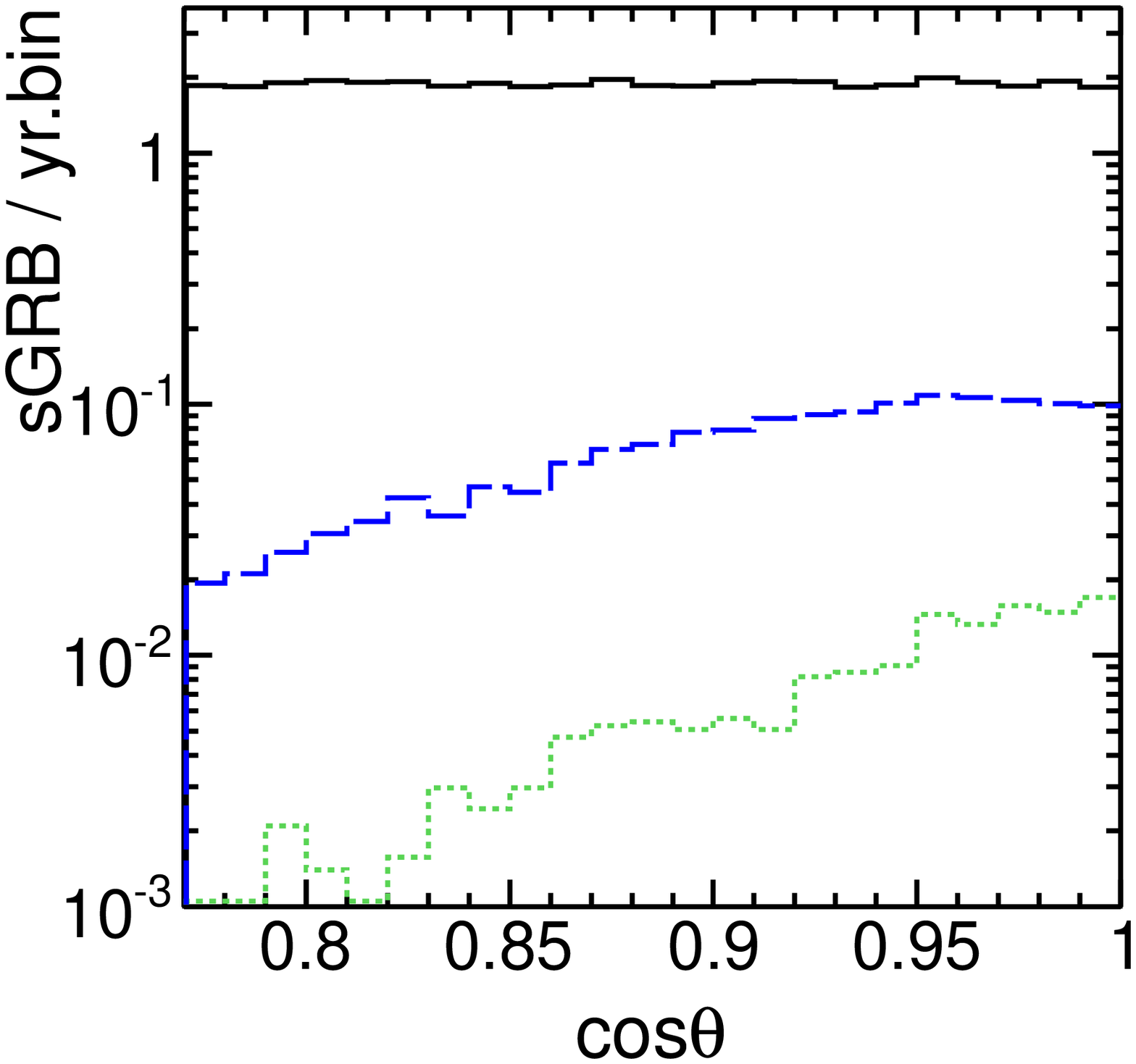}
  }
  \caption{\label{fig:sGRBresults} The three panels show the
    simulated distributions of short GRBs/yr detectable by Fermi GBM
    (black-solid lines) in fluence, redshift and zenith . The blue-dashed
    lines are GRBs detected by HAWC using the main DAQ, with default
    parameters. The green-dotted lines correspond the scaler DAQ.
}
\end{figure}

Using the model described above and using interpolations of HAWC's
effective area and parameterization of the background, we have
calculated the GRB detection
rate in HAWC. We find that sGRBs have a detection rate of 1.4 sGRB/yr (0.15 sGRB/yr) in the main DAQ (scaler DAQ). Figure \ref{fig:sGRBresults}
shows the distribution of sGRBs as detected by Fermi GBM as well as
HAWC with the main and scaler DAQs. We find that 90\% of the sGRBs
detected by the main DAQ (scaler DAQ) have a GBM fluence above $3.5
\times 10^{-7}$~erg$\cdot$cm$^{-2}$ ( $1.7 \times
10^{-6}$~erg$\cdot$cm$^{-2}$ ) . This corroborates that our result is
only modestly affected by the GBM trigger threshold. We also find that 90\% of the sGRBS detected by 
the main DAQ (scaler DAQ) have a redshift of $z\le 1.3$ ($z\le 1.4$). Finally 90\% of the sGRBs
detected by the main DAQ (scaler DAQ) have a zenith of $\theta
\le 28^\circ$ (31$^\circ$). We also find that all GRBs that can be
detected with HAWC scalers are detected by the main DAQ. 

\begin{figure}[ht]
  \centering
  \subfigure
  {  
    \includegraphics[width=.31\textwidth]{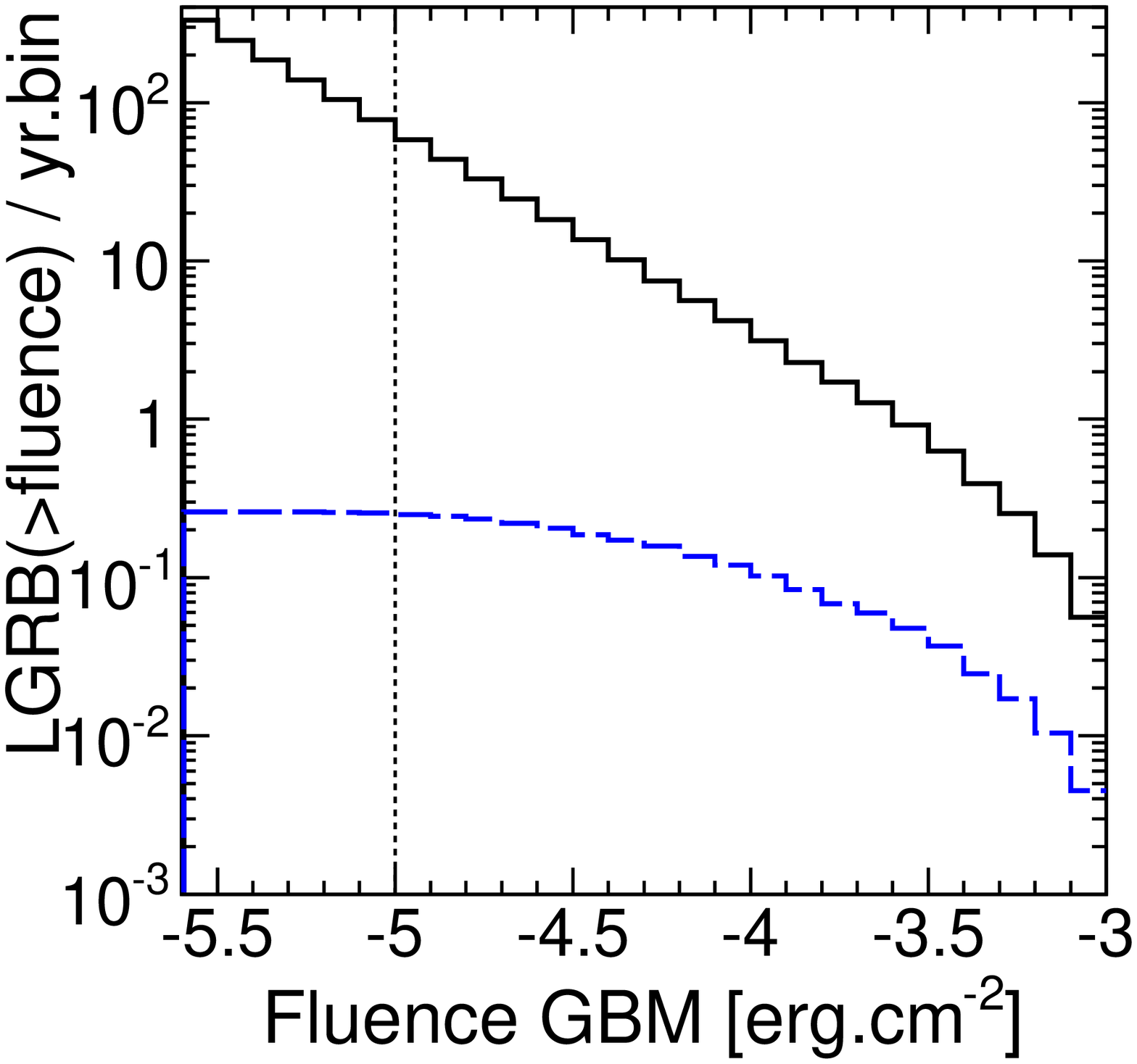}
  }
  \subfigure
  {
    \includegraphics[width=.31\textwidth]{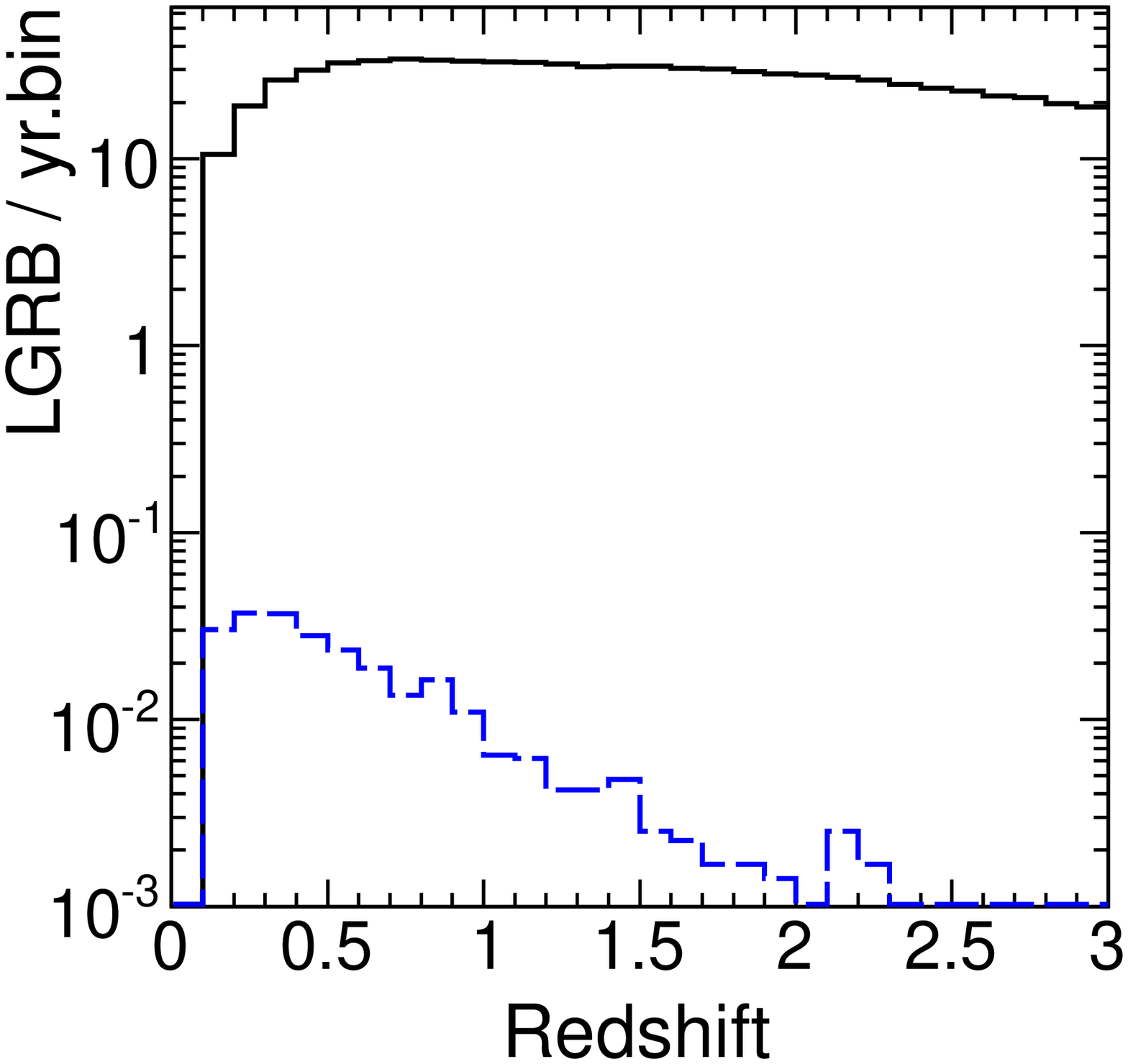}
  }
  \subfigure
  {
    \includegraphics[width=.31\textwidth]{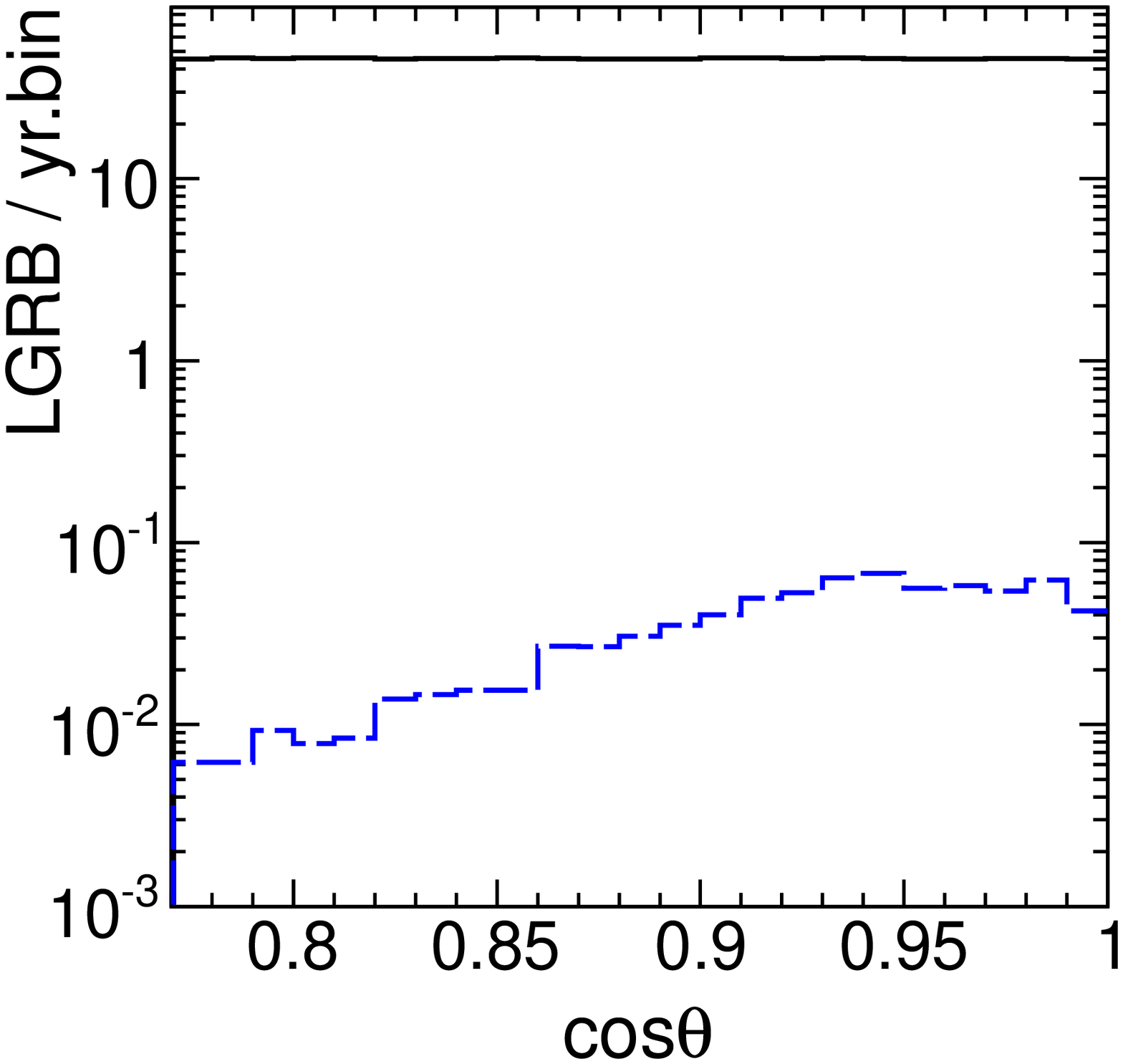}
  }
  \caption{\label{fig:LGRBresults} 
The three panels show the
    simulated distributions of long GRBs/yr detectable by Fermi GBM (black-solid lines) in
    fluence, redshift and zenith . The blue-dashed
    lines are GRBs detected by HAWC using the main DAQ, with default
    parameters. The green-dotted lines correspond the scaler DAQ.
}
\end{figure}

For LGRBs we find that the scaler DAQ has very poor prospects and we
can't calculate it reliably with the simulated statistics of
pseudo-GRBs, so we don't report the results here. For the main DAQ
the detection rate of LGRBs is 0.25 GRB/yr. Figure \ref{fig:LGRBresults}
shows the distribution of LGRBs as detected by Fermi GBM as well as
HAWC with the main DAQ. We find that 90\% of the LGRBs detected by the
main DAQ have a GBM fluence above $7 \times 10^{-6}$
erg$\cdot$cm$^{-2}$. We also find that 90\% of the sGRBS detected by
the main DAQ have a redshift of $z\le 1.1$. Finally 90\% of the sGRBs 
detected by the main DAQ have an elevation of $\theta \le 32^\circ$.

For both types of GRBs we have estimated the energy range of photons
that contribute to $>5\sigma$ detection. For the main DAQ 90\% of the
photons detected from a typical GRB are in the $\approx$50-500~GeV. For
scalers the range is $\approx$6-200~GeV. This is a reflection of the
different energy sensitivity of each of the two systems. 

For both types of GRBs we find what is reasonably expected: bright
GBM GRBs, with small redshift and high elevation are the ones with
best prospect for detection by HAWC. 

Since the GRB spectrum may be cutoff  at energies that are
below the EBL cutoff, we have simulated a universal
cutoff for all GRBs in the GRB reference frame. We have used cutoffs
from 100~GeV to 500~GeV. Results are summarized in Table
\ref{tab:cutoff}. For both main DAQ and scaler DAQ, the rate is
reduced by half if the universal cutoff is somewhere between 200 and
300~GeV in the GRB reference frame.


\begin{center}
\begin{table}
  \begin{tabular}{| c | c | c | c  |} 
    \hline
    Cutoff & main DAQ sGRB /yr & Scaler sGRB /yr & main DAQ LGRB /yr \\ 
    \hline
    n/a & 1.4 & 0.15  & 0.25  \\ 
   500~GeV & 1.3  & 0.12 & 0.22 \\
   400~GeV & 1.2 & 0.11 & 0.20  \\
   300~GeV & 0.97 & 0.10  & 0.15 \\
   200~GeV & 0.54 & 0.07 & 0.08  \\
   150~GeV & 0.27 & 0.05 & 0.04 \\
   100~GeV & 0.07  & 0.02 & 0.01\\
   \hline
\end{tabular}
\caption{\label{tab:cutoff} GRB rate with HAWC as a function of a
  universal cutoff in the GRB reference frame.} 
\end{table}
\end{center}

\section{$>$10~GeV detection rate in Fermi LAT}
\label{sec:LAT}

To test whether the model describe here is reasonable, we have
tested it against Fermi LAT. We have used our model to calculate the
rate of GRBs with at least one photon above 10~GeV in Fermi LAT. We
assume that the astrophysical 
backgrounds at 10~GeV and higher are virtually zero. We use the
effective area at 10~GeV as a function of incidence angle that has been
published by the Fermi LAT collaboration \cite{LATperformance}. We also
assume the effective area is constant as a function of energy above
10~GeV for all incidence angles. 

We find a rate of sGRBs of 0.26 sGRB/year and an LGRB rate of 1.2
LGRBs/year. As of the time of writing, Fermi LAT has been in operation
for 4.7 years and has detected one sGRB above 10~GeV (GRB~090510), in
agreement with our model. In the same time period LAT has detected 6 LGRBs 
above 10~GeV, also in agreement with our calculation.

For sGRBs we find that LAT should detect $>10$~GeV photons for GBM fluence above
$5 \times 10^{-6} erg \cdot cm^{-2}$. For LGRBs, we find that LAT
should detect $>$10~GeV photons for a GBM fluence above $6 \times 10^{-5} erg \cdot
cm^{-2}$. Both these thresholds are not sharp, instead they depend on
the angle of incidence in Fermi LAT.

Our calculation of the rate of GRBs in Fermi LAT above 10~GeV is very simplistic. We ignore
multiple effects. For example, in normal operations Fermi LAT points away
from Earth, but if a GRB alert is issued by GBM, then
the satellite may be re-pointed so that Earth's limb could partially
fall in the field of view of LAT. This would tend to increase the rate
of GRBs seen above 10~GeV by LAT.

\section{Discussion}

Observations by Fermi LAT unequivocally show that GRBs produce
VHE photons. But the
observed sample of GRBs that produce high energy photons is very
limited and with a wide variety of characteristics. In trying to
describe the 1~GeV-1~TeV emission by GRBs, assumptions about spectral
shape must be made and the spectrum must be extrapolated in energy
over a few orders of magnitude. The calculations performed  here are
critically sensitive to the choice of extrapolation. A cautionary tale of predictions that
depend on extrapolations is the pre-launch calculation of the GRB rate in Fermi
LAT \cite{LATrate} which were overestimated. Nevertheless our
model is consistent with current observations, save of the fraction of
LGRBs that have an additional power law in the spectrum. We
corroborated our model with respect to GRB observations by Fermi LAT
and we found good agreement. If we assume that only one third of LGRBs
have an additional power law, then the total GRB rate in HAWC is reduced to
1.51 GRBs/yr. Because both sGRBs detected with high signal to noise
ratio by Fermi LAT have an additional power law, we don't have good
knowledge of the fraction of sGRBs that do have these power
law. It is a critical assumption of our model that additional hard
power laws a very common in short-hard GRBs.

We find that HAWC is most sensitive to short-hard GRBs. In our model
we assume that half the VHE fluence is in temporal coincidence with
Fermi GBM, from which it follows that HAWC is most sensitive in the
prompt phase of the GRB. Interestingly the rate of short-hard GRBs in
HAWC is higher than the rate in Fermi LAT above 10~GeV. Interestingly
the rate of short-hard GRBs in HAWC is higher than the rate in Fermi
LAT above 10~GeV. It is also worth noting that the subpopulation of
GRBs visible to HAWC is almost  entirely separate from the population
predicted to be accessible to CTA in \cite{CTAgrbRate}, as the latter will almost
exclusively detect LGRBs.  

The detection rate of GRBs in HAWC can be as high as 1.65
GRBs/year. The rate is lower if the spectrum has cutoffs at energies
lower than that expected due to extra-galactic background light. A
lower detection rate that the one presented here should allow HAWC to
constrain the high energy spectral index, search for high energy
cutoff and/or the fraction of GRBs with additional power law components.

\section*{Acknowledgments}
IT acknowledges support by NSF grant PHY-1205807. RCG acknowledges
support from a Fermi Guest Investigator Grant.

\section*{Appendix - Short-Hard GRBs.}
\label{app:sGRBs}

Table \ref{tab:sGRB} has been compiled from Swift's database
\cite{SwiftData} and from individual GCN circulars. At first list was
drafted by selecting GRBs with a duration $T_{90}<2$~s. The GCN and
refereed publications for each GRB was reviewed to exclude potential
long-soft GRBs with $T_{90}<2$~s. We also include two GRBs that have
the characteristics of short-hard bursts, even though they have a
duration of $\approx$3~s.


\begin{center}
\begin{table}
\begin{tabular}{ | c | c | c | c | }
\hline
 GRB & Redshift & GBM $T_{90}$ (s) & References\\
 \hline
 050509B & 0.2248 & 0.03 &\cite{grb050509bA,grb050509bB}  \\
 050709 & 0.16 & 0.22& \cite{grb050709A,grb050709B} \\
 050724 & 0.258 & 3. &  \cite{grb050724A,grb050724B} \\
 050813 & 1.8 & 0.6 & \cite{grb050813A,grb050813B} \\
 051221A & 0.5464 & 1.4 & \cite{grb051221aA,grb051221aB} \\
 060502B &  0.287 & 0.128 & \cite{grb060502bA,grb060502bB} \\
 061006 & 0.4377 & 0.42 & \cite{grb061006,berger2007}\\
 061201 & 0.111 & 2. & \cite{grb061201A,grb061201B} \\
 061217  & 0.827  & 0.4 & \cite{grb061217,berger2007} \\
 070429B & 0.902 & 0.5 & \cite{grb070429bA,grb070429bB} \\
 070714B & 0.904 & 3. & \cite{grb070714bA,grb070714bB}  \\
 070724A & 0.457 & 0.4 & \cite{grb070724aA,grb070724aB} \\
 070809 & 0.2187 & 2. &\cite{grb070809A,grb070809B} \\
 071227 & 0.384 &1.8 & \cite{grb071227A,grb071227B} \\
 080905 & 0.1218 & 0.6 & \cite{grb080905A,grb080905B} \\
 090510 & 0.903 & 0.5 & \cite{GCNgrb090510A,GCNgrb090510B} \\
 100117A & 0.92 & 0.4  &\cite{grb100117aA,grb100117aB}  \\
 100206A & 0.41  & 0.13 & \cite{grb100206aA,grb100206aB}  \\
 101219A & 0.718 & 0.6 & \cite{grb101219aA,grb101219aB} \\
 \hline
\end{tabular}
 \caption{ \label{tab:sGRB} List of short-hard GRBs used to determine
   the redshift distribution}
\end{table}
\end{center}

\end{document}